\documentclass[final]{aipproc}

\layoutstyle{8x11single}

\usepackage{mathtools}
\usepackage{amsmath}
\usepackage{amsfonts}

\usepackage{color}

\begin{document}

\title{Cosmological solutions for a two-branes system in a vacuum bulk}


\author{Juan L. P\'erez}{
  address={Departamento de F\'isica, DCI, Campus Le\'on, Universidad
    de Guanajuato, 37150, Le\'on, Guanajuato, M\'exico.},
  ,email={einstein1_25@fisica.ugto.mx}
}

\author{Miguel A. Garc\'ia-Aspeitia}{
  address={Departamento de F\'isica, DCI, Campus Le\'on, Universidad
    de Guanajuato, 37150, Le\'on, Guanajuato, M\'exico.},
  ,email={aspeitia@fisica.ugto.mx}
}

\author{L. Arturo Ure\~na-L\'opez}{
  address={Departamento de F\'isica, DCI, Campus Le\'on, Universidad
    de Guanajuato, 37150, Le\'on, Guanajuato, M\'exico.},
  ,email={lurena@fisica.ugto.mx}
}

\author{Rub\'en Cordero.}{
  address={Escuela Superior de F\'isica y Matem\'aticas, IPN, M\'exico D. F.}
}

\begin{abstract}
We study the cosmology for a two branes model in a space-time of five
dimensions where the extra coordinate is compactified on an
orbifold. The hidden brane is filled with a real scalar field endowed
with a quadratic potential that behaves as primordial dark matter
field. This case is analyzed when the radion effects are negligible in
comparison with the density energy; all possible solutions are found
by means of a dynamical system approach.
\end{abstract}

\maketitle

\section{Introduction.}

Recent research in string theory and its generalization M-theory
\cite{Schwarz:2008kd,cuerdas1,cuerdas2,cuerdas3} have suggested that
11 dimensions are necessary to have a consistent quantum  string
theory. Inherited in these models are p-branes ($0<p<9$) and D-branes,
which would be the fundamental constituents of the Universe, as they
open the possibility that our visible universe can be a very large
D-brane extending over 3 spatial dimensions (see \cite{Bachas:1998rg} for a pedagogical review) 

The main idea of the string theory is that the Standard Model of
Particles Physics (SMPP) is made of open strings that are confined on
a D-brane, while gravity and other exotic fields, such as the dilaton
field, can freely propagate in the extra dimensions (the bulk). This
scenario is called brane cosmology, or brane-world cosmology, for
which the reduction to 5D of M-theory was first suggested by Horava
and Witten \cite{HW,300,301,302}, providing then the basis for many
brane models: Arkani-Dimopoulos-Dvali (ADD) \cite{ADD-1},
Randall-Sundrum type 1 (RS2), type 2 (RS2)\cite{randall,randall2}, and
Dvali-Gabadadze-Porrati (DGP) brane models of 5D gravity \cite{Dvali:2000hr}, among others.

In this paper, using the assumptions of RS1 models and some previous
results \cite{perez1,perez2}, we focus our attention in generalizations
of the solutions found in \cite{Binetruy:1999hy} for a vacuum 5D bulk. This
formalism generates a dynamical equation for the Hubble parameter in
our brane, $H_{c}$, which is closely related with the dynamics of the
Hubble parameter in hidden brane, $H_{0}$ \cite{Binetruy:1999ut};
\emph{i.e.}, the fields immersed in the hidden brane have a
gravitational influence on the dynamics in the visible brane.

We investigate the behavior of quadratic scalar field at early times
with the use of the dynamical brane equations. The main question we
want to address is: can the scalar field provide the source of
inflation at high energies through brane world dynamics? The behaviour
of the field is analyzed by means of dynamical system theory, under
the assumption that the radion has negligible effects on the brane
equations of motion. In the following we use units in which
$c=\hbar=1$.

\section{Two branes embedded in a 5D Space-time.} \label{Sec-2}

To start with, the universe will be modeled as a two brane system
embedded in a $5$-dimensional manifold. The fifth extra dimension is
represented by the coordinate $y$, and then the branes are located at
$y=0$ (visible), and $y=y_{c}$ (hidden), respectively, and their
corresponding matter fields are confined into $4$-dimensional
hypersurfaces. We write the most general metric in the form
\begin{equation}
  ds^{2} = - n^{2}(t,|y|) dt^{2} + a^{2}(t,|y|) g_{i j} dx^{i} dx^{j}
  +b^{2}(t,|y|) dy^{2}, \,  \label{metrica}
\end{equation}
where we have assumed that the two branes are dominated by perfect
fluid components which satisfy the following equations of state:
$p_{0}=\omega_{0}\rho_{0}$, and $p_{c}=\omega_{c}\rho_{c}$,
respectively. The energy momentum tensor is written as
\begin{equation}\label{tme} 
\tilde{T}^{A}_{B}=\hat{T}^{A}_{B}+\frac{\delta(y)}{b_{0}}diag(-\rho_{0},p_{0},p_{0},p_{0},0)+\frac{\delta(y-y_{0})}{b_{c}}diag(-\rho_{c},p_{c},p_{c},p_{c},0),
\end{equation}
where the first term corresponds to the bulk contribution and the second and third term corresponds to the branes embedded in the 5D manifold. As usual, the term $\hat{T}^{A}_{B}$ is in the form of a five dimensional cosmological constant, namely
\begin{equation}\label{Lambda-5}
\hat{T}_{AB}=-\frac{\Lambda_{5}}{\kappa^{2}_{(5)}}g_{AB},
\end{equation}

We have shown previously \cite{perez2} that the general metric that satisfies the $5$-dim Einstein's equations
($G_{_{AB}}=\kappa^{2}_{_{(5)}}T_{_{AB}}$) in a vacuum bulk is
\begin{eqnarray}
  \label{solucion1}
  a(t,y)&=&a_{0}\left[1+(m-1)\frac{\kappa^{2}_{(5)}}{6}\rho_{0}b_{0}y\right]^{1/(1-m)}, \\	
  \label{solucion2}	
  n(t,y)&=&n_{0}\left[1+\left(\frac{m}{2}+2+3\omega_{0}\right)\frac{\kappa^{2}_{(5)}}{6}\rho_{0}b_{0}y\right]\left[1+(m-1)\frac{\kappa^{2}_{(5)}}{6}\rho_{0}b_{0}y\right]^{m/(2-2m)}, \label{solucion2} \\
  \label{solucion3}
  b(t,y)&=&b_{0}\left[1+(m-1)\frac{\kappa^{2}_{(5)}}{6}\rho_{0}b_{0}y\right]^{m/(2-2m)},
\end{eqnarray}
where $\kappa^{2}_{(5)}$ is the $5$-dim gravitational constant related with the brane tension $\lambda_{0}$, and to the $4$- dim gravitational  constant as $\kappa^{4}_{(5)}=6\kappa^{2}_{(4)}/\lambda_{0}$. The functions $a_{0}$, $n_{0}$ and $b_{0}$ correspond to the time-dependent values of the metric coefficients in the brane at $y=0$, and 
$m$ is a parameter which determinate the bulk geometry. Substituting Eqs.~(\ref{solucion1}), (\ref{solucion2}), and~(\ref{solucion3}), into the boundary conditions,
\begin{equation}
\label{salto1}
\frac{\left[a'\right]_{0}}{a_{0}b_{0}}=-\frac{\kappa^{2}_{(5)}}{3}\rho_{0}, \ \ \ 
\frac{\left[a'\right]_{c}}{a_{c}b_{c}}=-\frac{\kappa^{2}_{(5)}}{3}\rho_{c} \, ,
\end{equation}
we can show that there exists a relationship among the evolution of energy densities, $\rho_c$
and $\rho_0$, in the two branes given by
\begin{equation}
  \label{conexion0}
  \rho_{c}=-\rho_{0}\left[1+(m-1)\frac{\kappa^{2}_{(5)}}{6}\rho_{0}b_{0}y_{c}\right]^{(m-2)/(2-2m)},\\
\end{equation}
where we have assumed that the metric coefficients satisfy the mirror symmetry, $\left[F^{\prime}\right]_{0}=2F^\prime |_{y=0+}$ and $\left[F^{\prime}\right]_{c}=-2F^\prime |_{y=y_{c}-}$, and a prime means derivative with respect to $y$. Similarly, using the boundary conditions, 
\begin{equation}
\label{salto2}
\frac{\left[n'\right]_{0}}{n_{0}b_{0}}=\frac{\kappa^{2}_{(5)}}{3}(3p_{0}+2\rho_{0}), \ \ \
\frac{\left[n'\right]_{c}}{n_{c}b_{c}}=\frac{\kappa^{2}_{(5)}}{3}(3p_{c}+2\rho_{c}),
\end{equation}
in Eqs.~(\ref{solucion1}), (\ref{solucion2}), and~\ref{solucion3}), we obtain that the equations of state (EoS), $\omega_c$ and $\omega_0$, are both related by the expresion
\begin{equation}  \label{conexion1}
  \omega_{c}=\frac{\omega_{0}+(\frac{m}{2}+2+3\omega_{0})(\frac{m}{6}-1)\frac{\kappa^{2}_{(5)}}{6}\rho_{0}b_{0}y_{c}}{1+(\frac{m}{2}+2+3\omega_{0})\frac{\kappa^{2}_{(5)}}{6}\rho_{0}b_{0}y_{c}}
  \, .
\end{equation}

Assuming a Friedmann-Robertson-Walker (FRW) metric on the visible
brane with $n_{c}=1$, we can write the Hubble parameters,
$H_{0}=-\epsilon\ \frac{\kappa^{2}_{(5)}}{6}n_{0}\rho_{0}$ and
$H_{c}=\epsilon\ \frac{\kappa^{2}_{(5)}}{6}n_{c}\rho_{c}$ in the two branes as \cite{perez2}
\begin{eqnarray}
  \label{hubble0-1}
  H_{0}&=&-\epsilon\
  \frac{\kappa^{2}_{(5)}}{6}\rho_{0}\frac{\left[1+(m-1)\frac{\kappa^{2}_{(5)}}{6}\rho_{0}b_{0}y_{c}\right]^{-m/(2-2m)}}{\left[1+\left(\frac{m}{2}+2+3\omega_{0}\right)\frac{\kappa^{2}_{(5)}}{6}\rho_{0}b_{0}y_{c}\right]}
  \, , \\
  \label{hubbleC-1}
  H_{c}&=&-\epsilon\
  \frac{\kappa^{2}_{(5)}}{6}\rho_{0}\left[1+(m-1)\frac{\kappa^{2}_{(5)}}{6}\rho_{0}b_{0}y_{c}\right]^{(m-2)/(2-2m)}
  \, ,
\end{eqnarray}
where $m\neq1$ is a parameter which defines the global geometry, and
$\epsilon=\pm1$. A complementary equation to solve the whole system is
the conservation equation in the hidden brane
\begin{equation}
\label{conservacion}
\dot{\rho}_{0}=-3(p_{0}+\rho_{0})H_{0}=-3\rho_{0}(1+\omega_{0})H_{0},
\end{equation}
where the sign of $\epsilon$ is chosen such that we obtain an
expanding universe within the visible brane. 

In the following sections, we consider the particular case $m=0$. In
this scenario, $b_{0}y_{c}=R$ is a constant and corresponds to a
stabilized radius of compactification; this is the most simple case in
the family of solutions found in \cite{perez2}.

\section{Scalar field dark matter component on the hidden brane.}

Scalar field dark matter (SFDM) is an interesting alternative model to
the dark matter problem, that may be able to resolve many conflicting
behavior of the standard model $\Lambda$-CDM in the formation of
structure at different scales. A very useful potential in SFDM
cosmology is the quadratic potential,
$V(\phi)=m_{\phi}^{2}\phi^{2}/2$, where $m_{\phi}$ is the scalar field
mass, whose value could be as low as $m_{\phi}\sim
10^{-22}eV$\cite{aspeita1,aspeita2}. Our scenario then considers a
SFDM dynamics at high energies using the braneworld context, and
assuming the topology shown in
Eqs.~(\ref{hubble0-1})-(\ref{hubbleC-1}). We investigate what kind of
conditions could make the SFDM be the source of inflation and the
possible consequences of it in physical observables, \emph{i.e.} the SFDM
coupled with brane gravity can modify the physical observables in inflation 
(e.g. spectral index, power spectrum, etc..) generating evidence of modified gravity
and likely that the SFDM could generate inflation in early universe ages.

Based in the results of the previous section, it is possible to compute that in the limit 
$\left|\frac{\kappa^{2}_{(5)}}{6} \rho_{0}R\right|\ll1$, Eqs. (\ref{hubble0-1}), and~(\ref{hubbleC-1}), can be written as
\begin{equation}\label{limite}
  H_{0}=H_{c}=-\epsilon\ \frac{\kappa^{2}_{(5)}}{6} \rho_{0}.
\end{equation}
Physically, this limit is consistent with an epoch in which the Hubble radius is much larger than  radius of compactification, namely $H_{0}^{-1}\gg R$. The above equation implies that both branes evolves closely related . If we assume
that $\rho_{0}>0$, then $\epsilon=-1$ in order to have positive solutions
for $H_{c}$. However, from Eq.~(\ref{conexion0}), we can see that the
energy densities in the branes have opposite signs due to the
$\mathcal{Z}_{2}$ symmetry imposed upon them, \emph{i.e.} the negative
effective energy density is a topological effect due to the mirror
symmetry as in the RS case.

Eq. (\ref{limite}) is consistent with the high energy limit in brane-world models, in which, traditionally the energy density is decomposed in two parts: $\rho_{0}=\rho_{m,0}+\lambda_{0}$. Substituting this last ansatz in Eq. (\ref{limite}) and squaring, we have \cite{Maartens:2003tw,Shiromizu:1999wj} 
\begin{equation}\label{hubble-ns}
H^2_{0}=\frac{\kappa^2_{(4)}}{3}\rho_{0,m}\left(1+\frac{\rho_{0,m}}{2\lambda_{0}}\right)+\frac{\kappa^4_{(5)}}{36}\lambda^2_{0},
\end{equation}
where, in the Randall-Sundrum approximation \cite{randall,randall2}, $\kappa^2_{(4)}\lambda_{0}+\Lambda_{5}=0$, and for our analysis, $\Lambda_{5}\approx 0$. In the high energy limit, $\rho_{0,m}\ll\lambda_{0}$, Eq. (\ref{hubble-ns}) becomes 
\begin{equation}\label{hubble-ns}
H^2_{0}\approx\frac{\kappa^2_{(4)}}{6}\frac{\rho^2_{0,m}}{\lambda_{0}}=\frac{\kappa^4_{(5)}}{36}\rho^2_{0,m}.
\end{equation} 
Now, we consider $\rho_{0}=\rho_{0,m}$ is composed only for a real scalar field living in the hidden brane, with
\begin{eqnarray}
  \rho_{0}=\frac{1}{2}\dot{\phi}^{2}+\frac{1}{2}m_{\phi}^{2}\phi^{2},
  \;\;\;
  p_{0}=\frac{1}{2}\dot{\phi}^{2}-\frac{1}{2}m_{\phi}^{2}\phi^{2} \, ,
\end{eqnarray}
and the equation of motion is the Klein-Gordon equation
\begin{equation}
  \ddot{\phi}+3H_{0}\dot{\phi}+m_{\phi}^{2}\phi=0.
\end{equation}

In order to analyze the dynamics of the system, it is convenient to
rewrite the equations of motion in terms of the new dimensionless
variables:
\begin{equation}
  x^{2}=\frac{\kappa^{2}_{(5)}}{12H_{0}}\dot{\phi}^{2} \, , \quad
  y^{2}=\frac{\kappa^{2}_{(5)}m_{\phi}^{2}}{12H_{0}}\phi^{2} \, ,
  \quad s=\frac{m_{\phi}}{H_{0}} \, ,
\end{equation}
where $s=m_{\phi}/H_{0}=m_{\phi}R_{H}$ is also directly proportional
to the Hubble radius. Then, the dynamical system of our system reads
\begin{eqnarray}
  x^{\prime}&=&-3\epsilon x^{3}-3x-sy, \nonumber \\
  y^{\prime}&=&-3\epsilon x^{2}y+sx, \label{sistema} \\
  s^{\prime}&=&-6\epsilon sx^{2},\nonumber
\end{eqnarray}
where a prime indicates derivative with respect to the $e$-foldings
number $N \equiv \ln a_{0}$. In order to have an expanding universe in
our brane, we choose $\epsilon=-1$

Variables $x$ and $y$ are subjected to the Friedmann constraint
$x^{2}+y^{2}= 1$, and then more appropriate variables are\cite{mayra}
\begin{equation}
  x=\cos \theta \, , \quad  y=\sin \theta \, .
\end{equation}
Then, the equations of motion~(\ref{sistema}) can be written as
\begin{eqnarray}\label{sistema2} 
  \theta^{\prime}&=&3\cos \theta \sin \theta + s \, , \nonumber\\
  s^{\prime}&=&6s \cos^2\theta \, .
\end{eqnarray}
The critical points, together with their stability, of this dynamical
system~(\ref{sistema2}), are listed in Table~\ref{tabla1}. The
stability of the critical points was determined by calculating the
eigenvalues and eigenvectors of the Jacobian matrix
\begin{equation}
  \mathcal{M}_{(\theta,s)}=\left[\begin{array}{cc}
      3\left(\text{cos}^2\theta-\text{sin}^2\theta\right) & 1 \\
      -12s\ \text{cos}\theta\text{sin}\theta  &
      6\text{cos}^{2}\theta\end{array}\right] \, .
\end{equation}

\begin{table}[ht!]
\caption{\label{tabla1} Properties of the critical points of the
  dynamical system~(\ref{sistema2}). All critical points are unstable,
but the saddle points are the source for inflationary solutions.}
\begin{tabular}{|c|c|c|}
\hline 
\hline
Critical point $\{\theta,s\}$, $n\in\mathbb{Z}$ & Eigenvalue& Stability  \\
\hline
\hline
$\{n\pi,0\}$ & $\{6,3\}$ & Unstable \\
\hline
$\{(n+\frac{1}{2})\pi,0\}$ & $\{-3,0\}$& Saddle\\
\hline \hline
\end{tabular}
\end{table}%

\begin{figure}[ht!]
\centering
\label{figure1}\caption{Analytical solutions of the dynamical
  system~(\ref{sistema2}). Note that the point $\{n\pi,0\}$ are
  unstable, whereas the points $\{(n+\frac{1}{2})\pi,0\}$ are
  saddle. We can see the expansion of the Universe starts in a kinetic
  dominated epoch, and goes to a potential dominated epoch, and
  the scalar field eventually ends up in an oscillatory behavior, see
  also Fig.~\ref{figure2}.}
\includegraphics[scale=1.1]{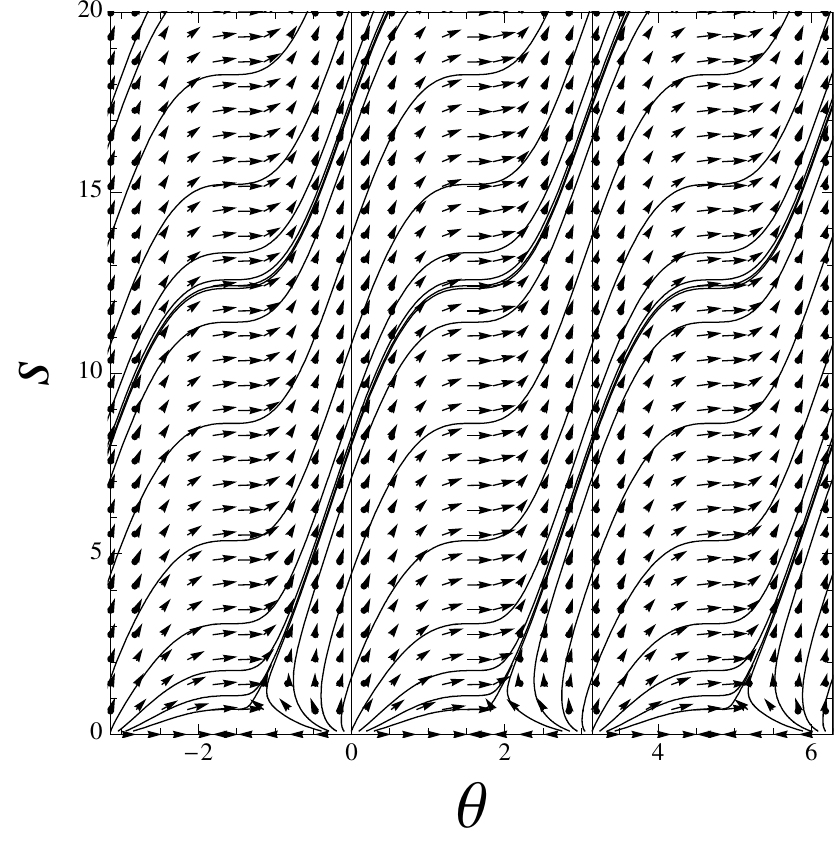}
\end{figure}

\begin{figure}[htbp]
\begin{tabular}{cc}
  \includegraphics[scale=0.72]{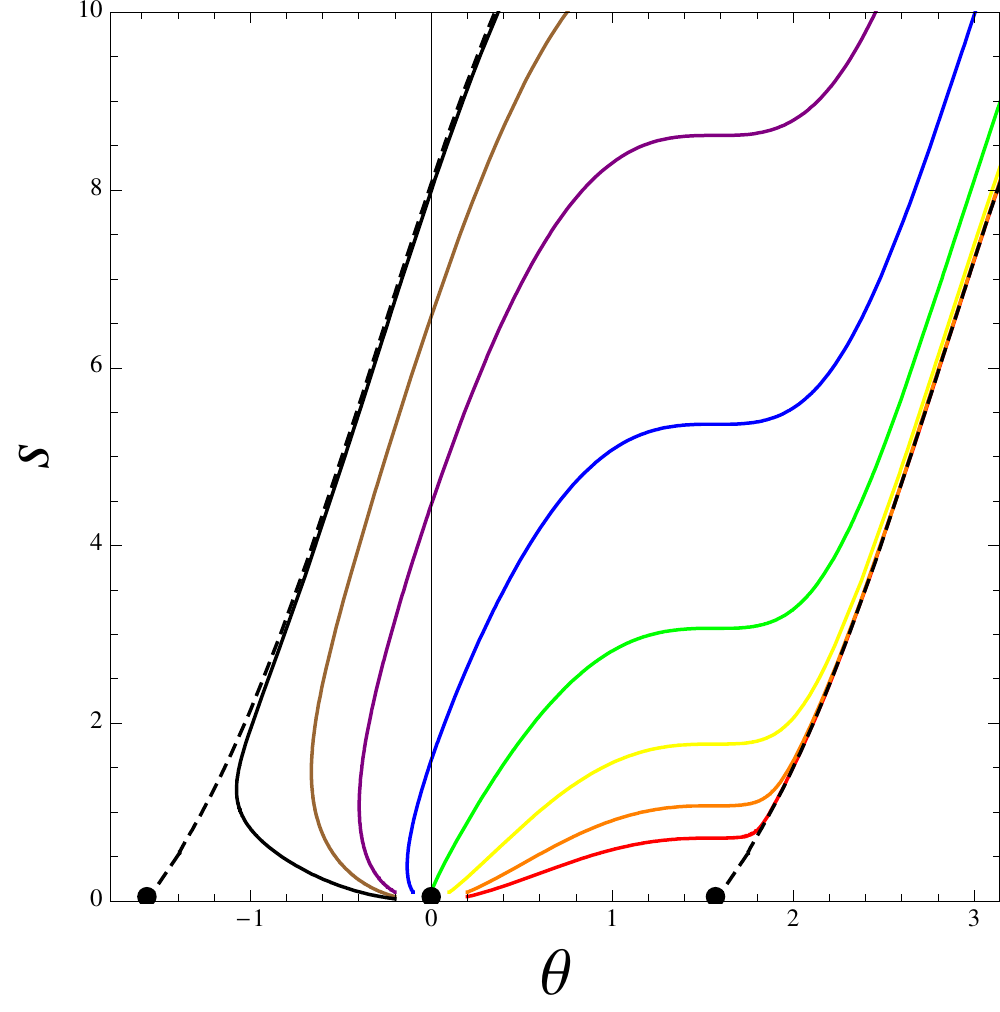} & \includegraphics[scale=0.9]{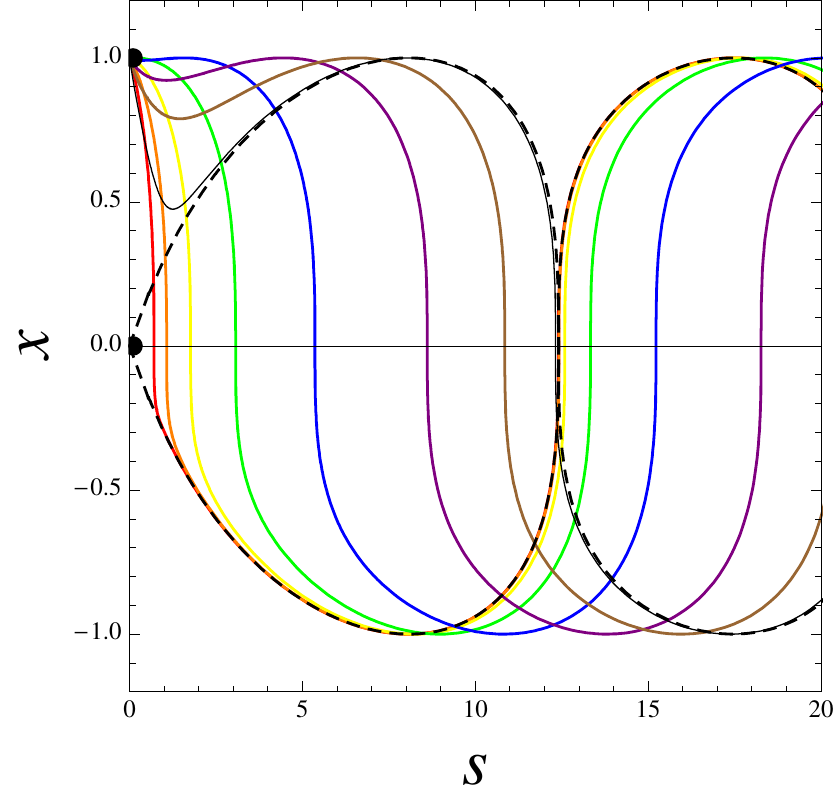}   \\
  &  \\
  \includegraphics[scale=0.9]{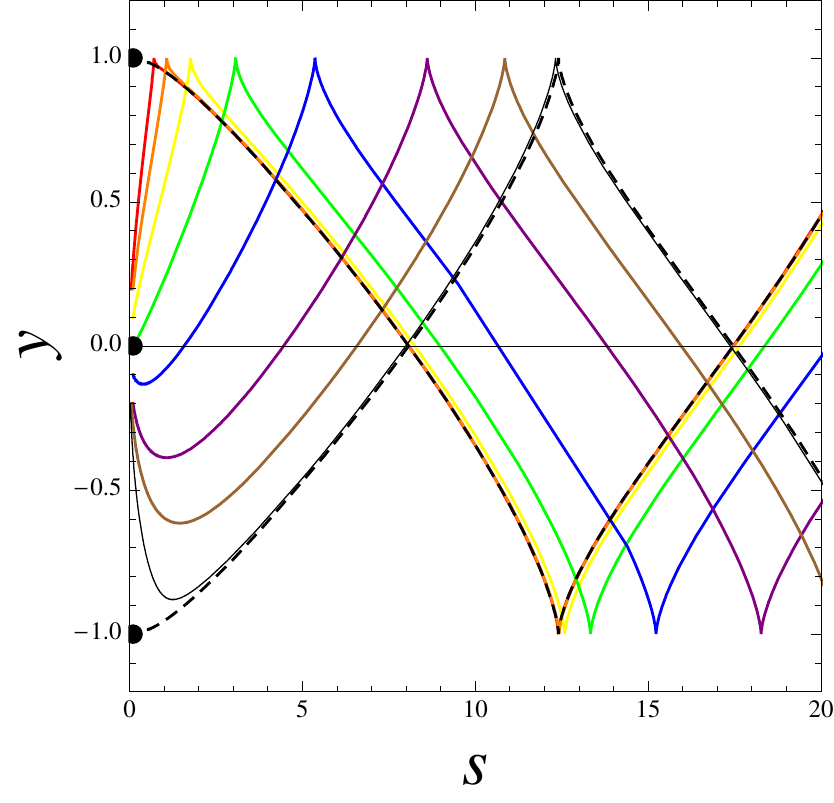}&  \includegraphics[scale=0.9]{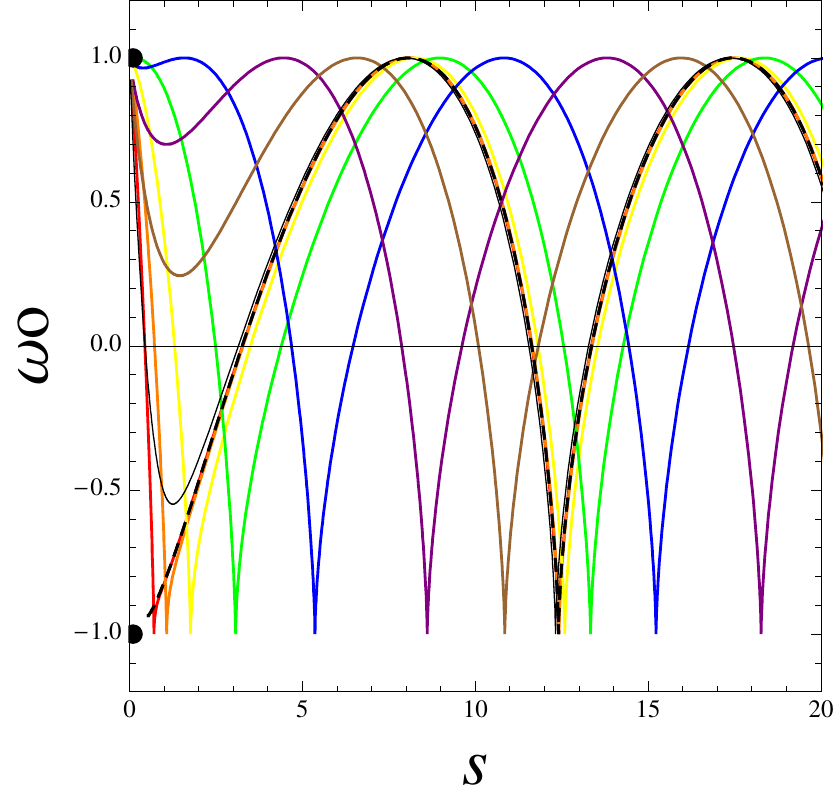}  
\end{tabular}
\caption{Evolution of the system (\ref{sistema2}) for different values
  of $\theta$ near the origin of coordinates (color lines); the
  universe begins at a kinetic dominated point. The inflationary
  solutions (dashed lines), which satisfy the slow-roll conditions,
  are located near of the saddle points. The corresponding evolutions
  of the kinetic $x$ and potential $y$ energies, and of EoS
  $\omega_0$, with respect to $s$ are also shown. We can notice the
  oscillatory behavior of all variables at late times.} 
\label{figure2}
\end{figure}

Typical solutions of the dynamical system on the $(\theta,s)$ plane
are shown in Fig.~\ref{figure1}. We observe that the points
$\{\theta=n\pi,s=0\}$ (where the kinetic energy is dominant) are
unstable. However, it is possible to see that the system eventually
evolves towards the critical point $\{\theta=\pi/2,s\neq0\}$ (where
the potential energy is dominant). Afterwards, the scalar field shows
an oscillatory behaviour at late times, this can be clearly seen in
Fig.~\ref{figure2}.

For sufficiently small values of variable $s$, the solutions move
closely to the saddle points, for which exists a natural exponential
expansion in our (visible) brane that satisfies the conditions for
inflation, see Fig.~\ref{figure2} and\cite{mayra}. It is possible to
observe that SFDM could be a good candidate not only for dark matter,
but the addition of brane world can likewise make it a strong candidate for
inflation.

\section{Conclusions} \label{concl}
Using previous results, for which the branes are connected by imposing
topological constraints and assuming a vacuum bulk, (with the aim of
obtain an exact solution for the metric coefficients), we study the
behavior of SFDM in the hidden brane and their visible effects in our
brane in high energies. Under the consideration that the radion
effects are negligible, we analyse the general solutions of the
equations of motion. The main results can be enumerated in the
following way:

\begin{itemize}
\item The universe goes towards different epochs: the first one is
  dominated by the kinetic energy of the field, the second one is
  dominated by the potential energy, and the last one is oscillatory.

\item The solutions that drive out the inflationary epoch, and that
  also satisfy the slow roll conditions, are located near the saddle
  critical points. This is always the case if $m_{\phi}\ll H$, which
  generates an exponential expansion that is typical of inflationary
  models.

\item It is important to remark that it is possible to obtain an
  expanding solution in the visible brane, as long as the effective
  (or apparently) energy density $\rho_{c}$ is negative. But, it is
  possible to argue that this is a consequence of the
  $\mathcal{Z}_{2}$ symmetry, as is also the case of RS models. 
\end{itemize}

A more general study should include an analysis of the evolution of
the Universe for both the high and low energy regimes. This is work
currently under preparation that will be published elsewhere.

\begin{theacknowledgments}
We are grateful to the Divisi\'on de Gravitaci\'on y F\'isica
Matem\'atica (DGFM) for the opportunity to present this work at the IX
Mexican School DGFM-SMF. This work was partially supported by PIFI,
PROMEP, DAIP-UG, CAIP-UG, CONACyT M\'exico under grant 167335 and
CONACyT postdoctoral grant. Instituto Avanzado de Cosmolog\'ia (IAC)
collaboration and Beyond Standard Theory Group (BeST) collaboration.
\end{theacknowledgments} 

\bibliographystyle{aipproc} 
\bibliography{librero1}

\end{document}